%% file: main_arxiv.tex
\documentclass[journal]{IEEEtran}

\usepackage[pdftex]{graphicx}
\usepackage[caption=false,font=footnotesize]{subfig}
\usepackage{xcolor}
\usepackage{url}
\usepackage[acronym]{glossaries}
\input{acronyms.tex}

\usepackage{makecell}
\newcommand{\tabitem}{~~\llap{\textbullet}~~}

\usepackage{tikz}
\usepackage{textcomp}
\usepackage{hyperref}
\usepackage{lipsum}

\newcommand\copyrighttext{%
  \footnotesize \textcopyright This work has been accepted for publication in the IEEE Wireless Communications. Copyright with IEEE. For more details, see the IEEE Copyright Policy.}
\newcommand\copyrightnotice{%
\begin{tikzpicture}[remember picture,overlay]
\node[anchor=south,yshift=10pt] at (current page.south) {\fbox{\parbox{\dimexpr\textwidth-\fboxsep-\fboxrule\relax}{\copyrighttext}}};
\end{tikzpicture}%
}

\begin{document}

\title{Use Cases for Terahertz Communications:\\An Industrial Perspective}

\author{Tommaso Zugno$^1$, Cristina Ciochina$^2$, Sharad Sambhwani$^3$,	Patrick Svedman$^4$, Luis M. Pessoa$^5$, Ben Chen$^6$,\\Per Hjalmar Lehne$^7$, Mate Boban$^1$, Thomas Kürner$^8$\\
\footnotesize{$^1$Huawei Technologies, Munich Research Center, Germany},
\footnotesize{$^2$Mitsubishi Electric R$\&$D Centre Europe (MERCE), France},
\footnotesize{$^3$Apple, Inc.},\\
\footnotesize{$^4$InterDigital Communications, Inc.},
\footnotesize{$^5$INESC TEC and Faculty of Engineering, University of Porto, Portugal},
\footnotesize{$^6$Beijing Jiaotong University, Beijing},\\
\footnotesize{$^7$Telenor Research \& Innovation, Fornebu, Norway},
\footnotesize{$^8$Technische Universität Braunschweig, Germany}
}

\maketitle
\copyrightnotice

\begin{abstract}
Thanks to the vast amount of available resources and unique propagation properties, \gls{thz} frequency bands are viewed as a key enabler for achieving ultrahigh communication performance and precise sensing capabilities in future wireless systems. 
Recently, the \gls{etsi} initiated an \gls{isg} on \gls{thz} which aims at establishing the technical foundation for subsequent standardization of this technology, which is pivotal for its successful integration into future networks.
Starting from the work recently finalized within this group, this paper provides an industrial perspective on potential use cases and frequency bands of interest for \gls{thz} communication systems.
We first identify promising frequency bands in the 100 GHz - 1 THz range, offering over 500~GHz of available spectrum that can be exploited to unlock the full potential of \gls{thz} communications.
Then, we present key use cases and application areas for \gls{thz} communications, emphasizing the role of this technology and its advantages over other frequency bands. 
We discuss their target requirements and show that some applications demand for multi-Tbps data rates, latency below 0.5~ms, and sensing accuracy down to 0.5~cm. 
Additionally, we identify the main deployment scenarios and outline other enabling technologies crucial for overcoming the challenges faced by \gls{thz} system.
Finally, we summarize the past and ongoing standardization efforts focusing on \gls{thz} communications, while also providing an outlook towards the inclusion of this technology as an integral part of the future \gls{6g} and beyond communication networks.
\end{abstract}


\glsresetall
\section{Introduction} 
The next generation of wireless communications technologies, i.e., the \gls{6g}, is expected to bring the digital and physical worlds even closer to each other across all dimensions, providing users with a holographic, haptic, and multi-sense experience. Turning this vision into reality requires not only enhancing the communication performance, but also delivering functionalities that are not currently supported.

One promising direction to meet this demand comes from the exploitation of new spectrum. So far, the frequency range between 0.1 and 10 \gls{thz}, also referred to as \gls{thz} spectrum, has remained underutilized, in part due to the lack of devices able to operate at such frequencies. Difficulties encountered in the design and fabrication of \gls{thz} devices have hindered the exploitation of this range and resulted in the so-called \gls{thz} gap, a ``dead zone'' between microwave, millimeter wave, and infrared spectra. As a result of recent advancements in the design and fabrication of \gls{rf} components, this gap is finally closing and mass-produced devices able to operate at \gls{thz} frequencies are expected to hit the market in the near future.  

\begin{figure*}[t]
    \centering
    \includegraphics[width=\linewidth]{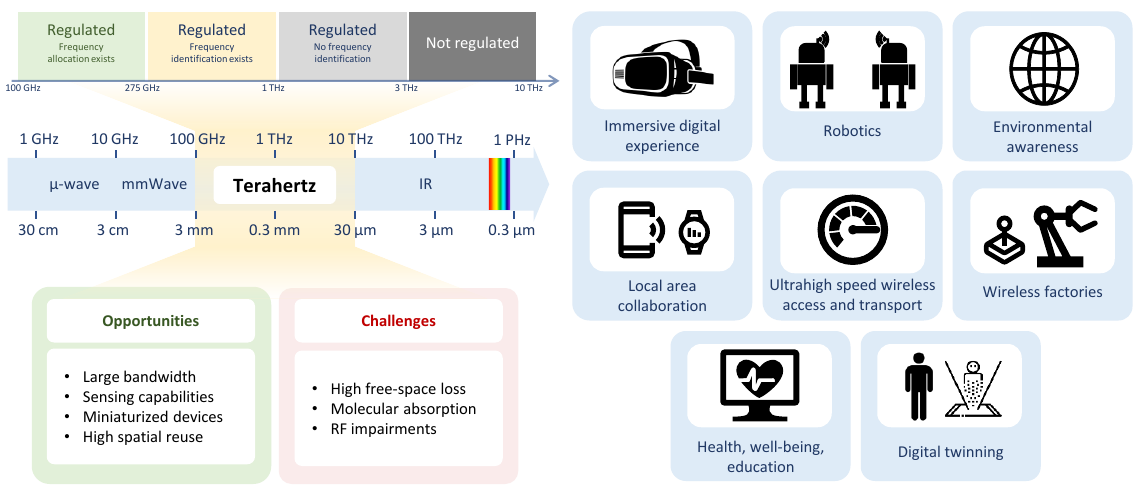}
    \caption{Overview of \gls{thz} communications and identified application areas.}
    \label{fig:thzspectrum}
\end{figure*}

Thanks to the large amount of available bandwidth, \gls{thz} systems can achieve higher capacity and lower latency than traditional wireless technologies, making them suitable for applications requiring high-speed and low-latency wireless access.  
The enhanced directionality of \gls{thz} systems contributes to reducing interference and maximizing spatial reuse, which is beneficial for applications with a high density of users.
Additionally, the small size of \gls{thz} antennas allows for the development of extremely compact transceivers for small-scale applications.
Along with ultra-high communication performance, \gls{thz} bands offer enhanced sensing capabilities and thus are particularly suitable for applications targeting quality-of-life goals such as advanced e-health, well-being, immersive entertainment, access to quality education for everyone everywhere, public safety, and safe work environments. 
Thanks to this great potential, \gls{thz} communications are considered by many as one of the main technological enablers for the next generations of wireless systems~\cite{Tong_Zhu_2021}. However, several challenges must be faced to fully exploit their capabilities, mostly due to the harsh and variable propagation conditions experienced at such high frequencies~\cite{THZBook}.

Several initiatives aiming at the development, standardization, and commercialization of this technology have been launched in recent years. In this regard, \gls{etsi} recently established an \gls{isg} focusing on \gls{thz} communications (\gls{etsi} \gls{isg} \gls{thz}), with the goal of providing appropriate models for \gls{thz} communications, thus laying the technical foundation for the development and standardization of the technology. The group currently brings together 63 industrial and academic institutions, including vendors, operators, universities, and research centers with expertise in relevant areas. 

In this paper, we provide a thorough analysis of use cases for \gls{thz} communication systems. We first summarize the outcomes that have been achieved so far by the \gls{etsi} \gls{isg} \gls{thz}, with a focus on potential use cases and frequency bands of interest.  
In particular, we provide a general overview of wireless communications at \gls{thz} frequencies, describe the current regulatory situation of the \gls{thz} spectrum, and introduce standardization and pre-standardization efforts focusing on this technology. 
Then, we present 19 use cases for \gls{thz} communications that have been identified by the \gls{etsi} \gls{isg} \gls{thz} and describe related application areas. 
Moreover, we provide further considerations on related aspects that are important for the development and standardization of \gls{thz} systems, 
including propagation environments and deployment scenarios, other enabling technologies, and potential requirements and \glspl{kpi}.

The rest of the paper is structured as follows. 
Section~\ref{sec:thz_land} provides an overview of \gls{thz} communications,
Section~\ref{sec:use_cases} presents the use cases that have been identified and describes the related application areas,
Section~\ref{sec:further_cons} discusses relevant aspects related to the identified use cases, and Section~\ref{sec:concl} concludes the paper.

\section{\gls{thz} communications and sensing landscape}\label{sec:thz_land}
This section overviews the regulatory situation of \gls{thz} frequency bands, describes the distinctive features of \gls{thz} signal propagation, and summarizes related standardization and pre-standardization efforts.

\subsection{\gls{thz} frequency bands}
The frequency range between 100~GHz and 10~\gls{thz} is referred to as the ``\gls{thz} range". 
The corresponding wavelengths are from 0.03~mm to 3~mm. Below this range, the \gls{mmwave} and microwave frequency ranges are found, already heavily utilized for communications and non-communications applications. Above 10~THz, the near- and mid-infrared spectrum starts.
Based on the regulatory status and application scenarios, the frequency band 100~GHz -- 10~THz can be divided in three ranges, as shown in Fig.~\ref{fig:thzspectrum}.

Frequency bands between 100 and 275~GHz are already allocated for terrestrial services on an international level through the \gls{itu} Radio Regulations~\cite{itu2020}, and up to 252 GHz, several bands allocated for passive services are explicitly protected against interference, typically for \gls{eess} and \gls{ras}. 
Already at \gls{wrc} 2019, 137~GHz of spectrum from 275 to 450 GHz were identified for the use by mobile and fixed services, i.e., \gls{thz} communications. This identification followed extensive sharing studies between \gls{thz} communications and passive services like \gls{eess} and \gls{ras}, which have already used this frequency band and need to be protected from harmful interference~\cite {itur2019}.
As technology evolves, the exploitation of \gls{thz} bands is becoming more feasible, and recent studies suggest that technological means can make sharing possible also in the protected bands~\cite{Pole23}, thus enabling the coexistence of passive, such as climate and weather monitoring, and active applications, such as mobile communications and radars. 
Further resolutions were approved at the \gls{wrc}~2023, ensuring contiguous allocations for fixed and mobile services between 232 and 239~GHz~\cite{wrc23}.

The \gls{etsi} \gls{isg} \gls{thz} identified a number of frequency bands as interesting for \gls{thz} communications~\cite{etsigr002}, which are reported in Tab.~\ref{tab:freqbands}. Between 100 and 275~GHz, 8 bands with sufficient contiguous bandwidth are allocated to fixed or mobile services on a co-primary basis, constituting 91.2~GHz of total available bandwidth. Above 275~GHz, 12 interesting bands have been identified for \gls{thz} communications purposes based on a combination of regulatory status and favorable propagation conditions, resulting in 488~GHz of bandwidth. 91~GHz of this range is identified for fixed and mobile services.

Further regulatory efforts are essential for the successful exploitation of the THz spectrum. One of the main issues is the development of regulatory frameworks on regional and national levels to support the intended applications. 
Another open problem concerns \gls{emf} exposure regulations, as there are currently no globally standardized guidelines specifically tailored for the \gls{thz} frequency range.
Finally, the design of effective sharing and protective methods is a key challenge to be addressed.

\begin{table}[]
\caption{Frequency bands of interest and regulatory status related to fixed or mobile services.}
\label{tab:freqbands}
\centering
\begin{tabular}{|c|c|c|}
\hline
\makecell[c]{\textbf{Band $[$GHz$]$}} & \makecell[c]{\textbf{Bandwidth $[$GHz$]$}} & \textbf{Regulatory status} \\
\hline
102   - 109.5 & 7.5 & Allocated\\
141   - 148.5 & 7.5 & Allocated \\
151.5 - 164   & 12.5 & Allocated\\
167   - 174.8 & 7.8 & Allocated\\ 
191.8 - 200   & 8.2  & Allocated\\
209   - 226   & 17   & Allocated\\
231.5 - 239.2 & 7.7 & Allocated\\ 
252   - 275   & 23  & Allocated\\ 
275   - 296   & 21  & Identified\\ 
296   - 306   & 10  & Not identified\\ 
 306 - 313 & 7   &  Identified\\
 313 - 318 & 5   &  Not identified\\
 318 - 321 & 3   &  Identified\\
 327 - 333 & 6   &  Identified\\
 333 - 356 & 23  &  Not identified\\
 356 - 368 & 12  &  Identified\\
 391 - 433 & 42  &  Identified\\
 452 - 520 & 68  &  Not identified\\
 598 - 722 & 124 &  Not identified\\
 786 - 953 & 167 & Not identified\\
\hline
\end{tabular}
\end{table}

\subsection{Distinct features of \gls{thz} communications}\label{sec:thz_features}
The large amount of radio resources available at \gls{thz} frequencies offer a promising solution to alleviate spectrum scarcity issues faced at lower frequency bands. The large bandwidth can be exploited to achieve higher capacity than traditional wireless systems, even with lower spectral efficiency, making \gls{thz} systems suitable for applications requiring high-speed wireless access.  
The large bandwidth also allows for a shorter slot duration and enables more frequent scheduling of users over time, thus reducing communication latency. 
Due to the short wavelength, \gls{thz} devices can embed a large number of antennas within a small form factor, achieving high gain and increased directionality.
This contributes to reduced interference and improved spatial reuse, which is beneficial for dense network deployments and applications with a high density of users. 
The small antenna size allows for the development of extremely compact transceivers, small enough to build wireless networks on chips and be implanted within the human body for advanced biomedical applications.

Additionally, \gls{thz} bands offer several advantages over lower frequencies for sensing applications, making \gls{thz} systems appealing for realizing the so-called \gls{isac} paradigm~\cite{Tong_Zhu_2021}. The larger available bandwidth can enhance the spatial resolution of sensing algorithms, allowing for more precise localization, detection, and imaging of objects~\cite{chen2022tutorial}. \gls{thz} antennas can achieve a higher directionality and lower sidelobes, improving the angular resolution and minimizing the occurrence of ghost targets. Given the short wavelength, \gls{thz} frequencies exhibit more pronounced scattering from surface roughness, which can be exploited to improve the detectability of objects. 
Moreover, unlike X and gamma rays, \gls{thz} radiation is non-ionizing and thus safe for human exposure. \gls{thz} waves can penetrate non-conductive materials such as plastic, paper, and textiles, making them suitable for detecting hidden objects or defects in security screening, quality control, and process monitoring applications.
Due to the peculiar behavior when interacting with various substances, \gls{thz} frequencies are commonly used for spectroscopy applications and for characterizing material properties~\cite{s19194203}. 

The potential of \gls{thz} communications has been proven by various experiments conducted in the past few years. For example, Northeastern University has demonstrated a multi-kilometer and multi-Gbps link operating at 210–240~GHz, 
while the EU-Japan ThoR project has demonstrated a 300~GHz bidirectional communication link achieving 20~Gbps over 160~m.

\begin{figure}[t]
    \centering
    \includegraphics[width=0.48\textwidth]{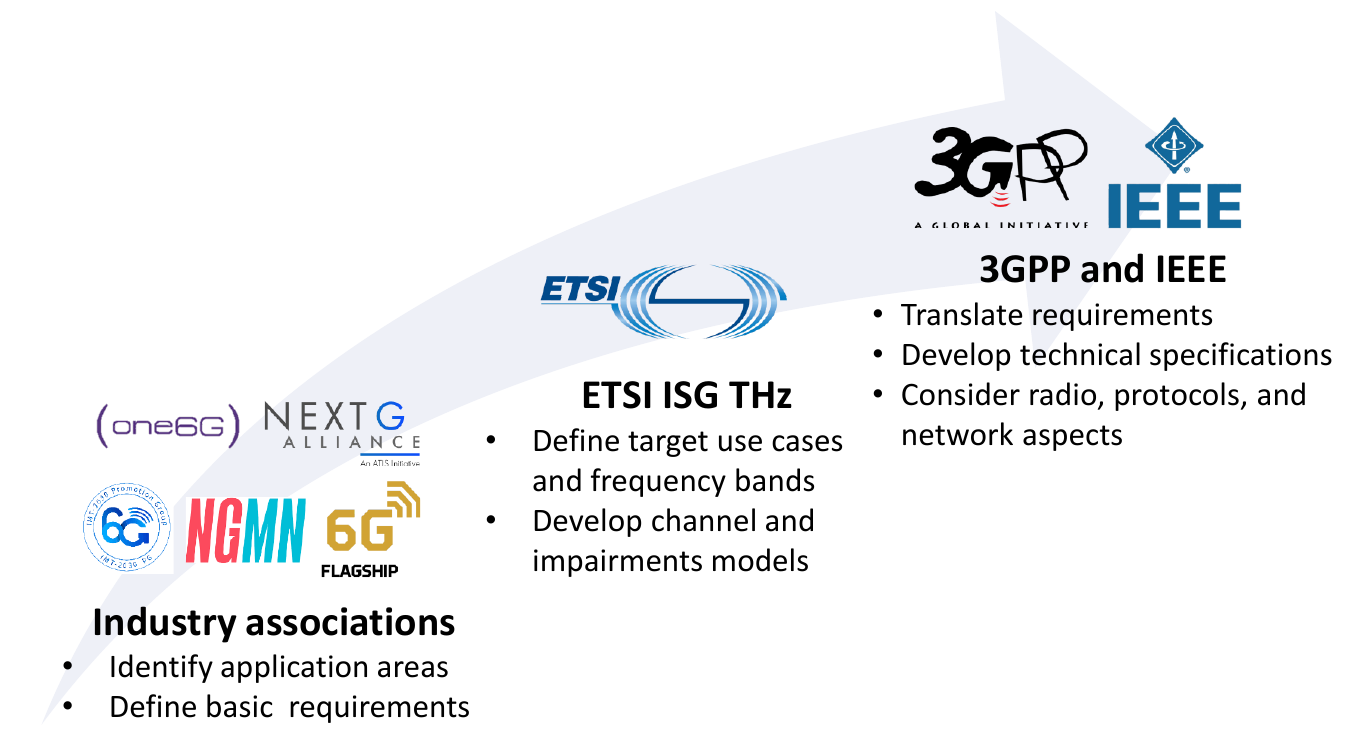}
    \caption{Standardization of \gls{thz} communication systems.}
    \label{fig:standardization}
\end{figure}

\subsection{Standardization and pre-standardization efforts}
Although \gls{thz} communications is a relatively new field, the first standardization activities started already more than 15 years ago. In 2008, the \gls{ieee} 802 commitee established a \gls{thz} Interest Group, which is now continuing its work as the \gls{ieee} 802.15 Standing Committee \gls{thz}. As a spin-off of these early activities, in 2017 the \gls{ieee} 802 completed the \gls{ieee} Std 802.15-3d-2017 as an amendment to \gls{ieee} Std 802.15.3-2016, which defines a new \gls{thz} physical layer for the frequency range 252-321 GHz featuring 8 different bandwidths between 2.16 GHz to 69.2 GHz. A revision published in 2023 \gls{ieee} Std 802.15.3-2023 extends the frequency range up to 450~GHz~\cite{802.15.3-2023}. 
Despite representing the first standard for THz communications, it is applicable to a limited set of use cases with fixed point-to-point links, and thus not suitable for mobile communication systems.

Major industrial associations for next-generation wireless systems foresee the inclusion of \gls{thz} communications in 6G standards and some of them are actively working on the development of this technology. For example, the European initiatives one6G and 6G FLAGSHIP, the North-American initiative Next G Alliance (NGA), the forum of world-leading mobile network operators Next Generation Mobile Networks (NGMN) Alliance, and the Chinese initiative IMT-2030 6G Promotion Group consider \gls{thz} communications as one of the key technologies for 6G and beyond. Bringing together multiple players in different domains and across different verticals, these associations aim to define the basic requirements and identify the application groups of interest for this technology.

Recently, \gls{etsi} has started a related pre-standardization activity by establishing the \gls{isg} \gls{thz} at the end of 2022. 
The work program of this group is divided into four work items covering different aspects of the \gls{thz} technology, including the identification of use cases and frequency bands of interest, channel measurements and modeling, and RF hardware modeling. In particular, work item DGR/\gls{thz}-001  studied the use cases that are relevant for \gls{thz} communication systems, identified possible deployment scenarios, and analyzed potential requirements and \glspl{kpi}. Similarly, work item DGR/\gls{thz}-002 investigated the frequency bands of interest, analyzed the current regulatory situation, and identified the incumbent services to be considered for coexistence studies. The main findings of these work items have been documented in publicly available reports, namely GR001~\cite{etsigr001} and GR002~\cite{etsigr002}. As future steps, work item DGR/\gls{thz}-003 will analyze specific radio propagation aspects for \gls{thz} communications, perform channel measurement campaigns for the identified use cases and frequency bands, and develop channel models for \gls{thz} bands. At the same time, DGR/\gls{thz}-004 will develop models and investigate possible approaches for the realization of \gls{thz} devices, circuits, and systems, characterize RF impairments to obtain suitable models in the \gls{thz} frequency range, and assess the overall device complexity, cost, and energy impact.

The final goal of the \gls{etsi} \gls{isg} \gls{thz} is to lay the foundation for the development and standardization of this technology. 
The outcomes of this activity are openly published group reports covering all the above-mentioned aspects, including channel and impairments models. 
This material will be available for subsequent standardization activities such as \gls{3gpp} and \gls{ieee}. 
An overview of different entities involved in the standardization process and their main roles is provided in Fig.~\ref{fig:standardization}.

\section{Use cases for \gls{thz} communication systems}\label{sec:use_cases}
The \gls{etsi} \gls{isg} \gls{thz} identified 19 potential use cases for \gls{thz} communication systems, which are listed in Table~\ref{tab:use_cases_1}-\ref{tab:use_cases_2} and documented in the group report GR001~\cite{etsigr001}.
From an application perspective, the identified use cases have been grouped into different areas. In the following, we describe the application areas and discuss the key aspects of the associated use cases.  
As shown in Table~\ref{tab:use_cases_1}-\ref{tab:use_cases_2}, the same use case may be relevant for multiple areas. 
Interested readers are referred to the group report for a more detailed description of each individual use case~\cite{etsigr001}.

\begin{table*}
    \centering
    \caption{Identified use cases.}
    \renewcommand{\arraystretch}{2}
    \begin{tabular}{p{0.16\textwidth}|p{0.25\textwidth}|p{0.2\textwidth}|p{0.3\textwidth}}
    \textbf{Use case} & \textbf{Application areas} & \textbf{Propagation environments} & \textbf{Other enabling technologies} \\
    \hline
    Remote surgery & 
    \makecell{\tabitem Immersive digital experience\\\tabitem Robotics\\\tabitem Health, Wellbeing, Education} & 
    \makecell{\tabitem On-body\\\tabitem Medical facility\\\tabitem Living room} &
    \makecell{\tabitem AI/ML\\\tabitem Advanced MIMO\\\tabitem Sensing, imaging, positioning\\\tabitem Energy efficient design for portable devices\\\tabitem Edge computing} \\ 
    \hline
    In airplane / train cabin entertainment & 
    \makecell{\tabitem Local area collaboration\\\tabitem Ultrahigh-speed wireless \\\hspace{0.3cm} access and transport \\\tabitem Health, wellbeing, education} & 
    \makecell{\tabitem Airplane cabin\\\tabitem Train cabin} &
    \makecell{\tabitem Advanced MIMO\\\tabitem Advanced relays\\\tabitem Sensing, imaging, positioning\\\tabitem Energy efficient design for portable devices\\\tabitem D2D, mesh networks\\\tabitem Edge computing} \\
    \hline
    Cooperative mobile robots & 
    \makecell{\tabitem Robotics\\\tabitem Local area collaboration\\\tabitem Wireless factories\\\tabitem Digital twinning} & 
    \makecell{\tabitem Factory} &
    \makecell{\tabitem RIS\\\tabitem Sensing, imaging, positioning\\\tabitem Energy efficient design for portable devices\\\tabitem D2D, mesh networks\\\tabitem Edge computing} \\
    \hline
        Hazardous material work & 
    \makecell{\tabitem Robotics\\\tabitem Wireless factories\\\tabitem Environmental awareness\\\tabitem Health, wellbeing, education} & 
    \makecell{\tabitem Factory} &
    \makecell{\tabitem Sensing, imaging, positioning\\\tabitem Energy efficient design for portable devices} \\
    \hline
        Remote education & 
    \makecell{\tabitem Immersive digital experience\\\tabitem Robotics\\\tabitem Ultrahigh-speed wireless \\\hspace{0.3cm} access and transport\\\tabitem Health, wellbeing, education\\\tabitem Digital twinning} & 
    \makecell{\tabitem Classroom} &
    \makecell{\tabitem AI/ML\\\tabitem Sensing, imaging, positioning\\\tabitem Energy efficient design for portable devices\\\tabitem Edge computing} \\
    \hline
        Wireless X-haul transport for fixed and mobile links & \makecell{\tabitem Ultrahigh-speed wireless \\\hspace{0.3cm} access and transport} &
    \makecell{\tabitem Open space\\\tabitem Highway\\\tabitem Urban street\\\tabitem Urban canyon\\\tabitem Stadium\\\tabitem Square} &
    \makecell{\tabitem Advanced MIMO} \\
    \hline
        Wireless data centres & 
    \makecell{\tabitem Ultrahigh-speed wireless \\\hspace{0.3cm} access and transport} & 
    \makecell{\tabitem Data centre} &
    \makecell{\tabitem Advanced MIMO\\\tabitem RIS} \\
    \hline
        Interactive immersive XR & 
    \makecell{\tabitem Immersive digital experience\\\tabitem Environmental awareness\\\tabitem Health, wellbeing, education\\\tabitem Digital twinning} & 
    \makecell{\tabitem Living room\\\tabitem Office} &
    \makecell{\tabitem Advanced MIMO\\\tabitem RIS\\\tabitem Advanced relays\\\tabitem Sensing, imaging, positioning\\\tabitem Energy efficient design for portable devices\\\tabitem D2D, mesh networks\\\tabitem Edge computing} \\
    \hline
        Mission critical Immersive digital experience & 
    \makecell{\tabitem Immersive digital experience\\\tabitem Robotics\\\tabitem Environmental awareness\\\tabitem Health, Wellbeing, Education} & 
    \makecell{\tabitem Medical facility\\\tabitem Living room\\\tabitem Factory\\\tabitem Classroom\\\tabitem Data centre\\\tabitem Office\\\tabitem Open space\\\tabitem Urban street\\\tabitem Urban canyon\\\tabitem Stadium\\\tabitem Square} &
    \makecell{\tabitem Sensing, imaging, positioning\\\tabitem Energy efficient design for portable devices\\\tabitem D2D, mesh networks\\\tabitem Edge computing} \\
    \hline
        Real-time industrial control & 
    \makecell{\tabitem Robotics\\\tabitem Environmental awareness\\\tabitem Wireless factories} & 
    \makecell{\tabitem Factory\\\tabitem Intra-machine/device\\\tabitem Inter-machine} &
    \makecell{\tabitem RIS\\\tabitem Sensing, imaging, positioning\\\tabitem Edge computing} \\
    \hline
        Simultaneous imaging, mapping, localization & 
    \makecell{\tabitem Robotics\\\tabitem Environmental awareness} & 
    \makecell{\tabitem Living room\\\tabitem Classroom\\\tabitem Hallway\\\tabitem Meeting room\\\tabitem Office\\\tabitem Stadium\\\tabitem Square} &
    \makecell{\tabitem AI/ML\\\tabitem Sensing, imaging, positioning\\\tabitem Energy efficient design for portable devices} \\
    \hline
        Commissioning of industrial plants & 
    \makecell{\tabitem Robotics\\\tabitem Wireless factories\\\tabitem Digital twinning} & 
    \makecell{\tabitem Factory} &
    \makecell{\tabitem AI/ML\\\tabitem Sensing, imaging, positioning\\\tabitem Edge computing} \\
    \hline
        Grand events with ultra-high throughput & 
    \makecell{\tabitem Immersive digital experience\\\tabitem Ultrahigh-speed wireless \\\hspace{0.3cm} access and transport} & 
    \makecell{\tabitem Stadium\\\tabitem Square} &
    \makecell{\tabitem Advanced MIMO\\\tabitem Energy efficient design for portable devices} \\
    \hline
\end{tabular}
    \label{tab:use_cases_1}
\end{table*}

\begin{table*}
    \centering
    \caption{Identified use cases (continued).}
    \renewcommand{\arraystretch}{2}
    \begin{tabular}{p{0.16\textwidth}|p{0.25\textwidth}|p{0.2\textwidth}|p{0.3\textwidth}}
    \textbf{Use case} & \textbf{Application areas} & \textbf{Propagation environments} & \textbf{Other enabling technologies} \\
    \hline
        Ultra-high throughput for indoor users & 
    \makecell{\tabitem Immersive digital experience\\\tabitem Local area collaboration\\\tabitem Ultrahigh-speed wireless \\\hspace{0.3cm} access and transport\\\tabitem Wireless factories} & 
    \makecell{\tabitem Living room\\\tabitem Classroom\\\tabitem Hallway\\\tabitem Meeting room\\\tabitem Office} &
    \makecell{\tabitem Advanced MIMO\\\tabitem Advanced relays\\\tabitem Energy efficient design for portable devices} \\
    \hline
        Intra-device communications & 
    \makecell{\tabitem Ultrahigh-speed wireless \\\hspace{0.3cm} access and transport} & 
    \makecell{\tabitem Intra-machine/device} &
    \makecell{\tabitem Energy efficient design for portable devices} \\
    \hline
        Local area collaboration for fixed or low mobility applications & 
    \makecell{\tabitem Local area collaboration\\\tabitem Ultrahigh-speed wireless \\\hspace{0.3cm} access and transport \\\tabitem Health, wellbeing, education} & 
    \makecell{\tabitem On-body\\\tabitem Medical facility\\\tabitem Living room\\\tabitem Factory	\\\tabitem Classroom\\\tabitem Hallway\\\tabitem Meeting room\\\tabitem Office\\\tabitem Open space\\\tabitem Stadium\\\tabitem Square} &
    \makecell{\tabitem Energy efficient design for portable devices\\\tabitem D2D, mesh networks} \\
    \hline
        Local area collaboration for vehicular applications & 
    \makecell{\tabitem Local area collaboration} & 
    \makecell{\tabitem Open space\\\tabitem Highway\\\tabitem Urban street\\\tabitem Urban canyon} &
    \makecell{\tabitem Energy efficient design for portable devices\\\tabitem D2D, mesh networks} \\
    \hline
        Predictive maintenance and diagnostics & 
    \makecell{\tabitem Robotics\\\tabitem Environmental awareness\\\tabitem Wireless factories\\\tabitem Digital twinnig} & 
    \makecell{\tabitem Factory} &
    \makecell{\tabitem AI/ML\\\tabitem Edge computing} \\\hline
\end{tabular}
    \label{tab:use_cases_2}
\end{table*}

\subsection{Robotics}
\gls{thz} links can provide the required data rates to enable efficient robot-to-network and robot-to-robot communications for a variety of robots, such as automated guided vehicles, industrial robots, drones, etc.
For example, the offloading of video or sensor data to a neighboring edge server for real-time processing requires high data rates (ranging between 600 Mbps and 3 Gbps depending on the video quality~\cite{etsigr001}), which can be provided through high-capacity \gls{thz} links. \gls{thz} bands are also suitable for establishing fast short-range connections among robots for data-sharing purposes.
Furthermore, \gls{thz} sensing can provide accurate robot positioning, as well as assist robots in achieving high-resolution environmental mapping, including information on material composition and surface textures of surrounding objects.

\subsection{Environmental Awareness}
As described in Sec.~\ref{sec:thz_features}, \gls{thz} bands provide unique sensing capabilities which can enable wireless systems to achieve a holistic view of the surroundings.
For example, unlike current deployments, outdoor \gls{thz} networks can monitor the emission of harmful gases and pollutants into the atmosphere by leveraging the unique molecular absorption profiles of \gls{thz} signals, thus contributing to environmental awareness and climate action goals. 
Additionally, \gls{thz} systems can be used to detect the presence of toxic substances and hazardous materials in workplaces and indoor environments, thereby improving situational awareness and human safety.

\subsection{Immersive Digital Experience}
Immersive digital experiences characterized by real-time interaction with other remote humans, avatars, and objects in virtual worlds can be provided using wearables (such as \gls{xr} headsets, haptic gloves, holographic displays). 
To enable a seamless user experience, ultimate \gls{xr} applications require extremely high data rates and low latency, which are beyond the capabilities of today’s \gls{mmwave} 5G systems, thus giving the reason for exploiting \gls{thz} bands. 
In this context, \gls{thz} links can be used to connect wearable devices to the network or to create side connections among devices worn by the same user or by nearby users. Moreover, real-time gesture and body pose detection, as well as precise localization services, can be assisted by the sensing capabilities provided by \gls{thz} signals.
\gls{thz}-enabled \gls{xr} can also be used in use cases with mission-critical constraints, such as supporting public safety personnel during their operations and assisting doctors in performing surgeries on remote patients.

\subsection{Local Area Collaboration}
 Thanks to the available spectrum and limited propagation range, utilizing \gls{thz} links for the purpose of transferring data between devices in close proximity can be beneficial.
 Compared to lower frequencies, \gls{thz} bands can enable a more efficient sharing of the spectrum and minimize the risk of collisions in the wireless channel.   
In this way, \gls{thz} can support many use cases for local collaboration among wireless terminals (e.g., smartphones, wirelessly connected watches and glasses, notebooks, tablets, or storage devices), such as fast media sharing and high-speed data backup. Also, since small wearables may have limited computational capabilities, \gls{thz} links can assist the offloading of compute tasks to higher-end devices within reach, such as smartphones, tablets, and notebooks. 

\subsection{Ultra-high-speed wireless access and transport}
Given the large bandwidth and the high spatial reuse, \gls{thz} networks can support a much higher connection density than lower frequency bands.
The dense deployment of \gls{thz} microcells can provide ultra-high-speed internet access in stadiums and festivals, enabling audiences to access and share high-resolution media content in real-time. 
In small indoor scenarios, such as train or plane cabins, \gls{thz} access points located above the seats can provide passengers with fast access to a local media server for streaming high-definition videos and movies, thus improving their travel experience.
In addition, \gls{thz} links can support wireless transport applications, such as wireless back/front/mid-hauling, able to provide higher capacity and reduced interference compared to solutions at lower frequency bands. 

\subsection{Wireless Factories}
Current industrial facilities are still largely based on wired communications technologies which present major issues, the main ones being increased costs, size, and weight due to cabling, and the need for special measures to connect devices in motion, such as reinforced cables, contact strips, etc. 
Wireless technologies based on \gls{thz} communications have the potential to meet the demanding requirements envisioned for future industrial applications. For example, they can support real-time information exchange between industrial machines and virtual controllers deployed at an edge server, thus enabling a more flexible management of the factory. 

\subsection{Health, well-being, education}
The \gls{thz} technology offers unique capabilities for supporting novel medical, healthcare, and educational applications.
For example, it can be used to build extremely small devices with communication and sensing capabilities that can be implanted in the body to monitor critical health parameters, such as blood glucose levels, heart rate, and oxygen saturation. These sensors can wirelessly transmit the collected data to user devices for continuous health monitoring. 
\gls{thz} access points installed in classrooms can provide fast wireless access to students equipped with wireless \gls{xr} devices and holographic displays for improving their learning experience. 
Moreover, \gls{thz} sensors integrated into student's smartphones can enable the characterization of their surroundings, such as analyzing biological samples, and deepen their understanding of natural sciences.

\subsection{Digital twinning}
\gls{thz} technologies offer enhanced support for digital twin applications.
\gls{thz} imaging techniques can be used to produce precise digital replicas of mechanical components and other physical objects, enabling the creation of accurate digital twins. The shorter wavelength compared to \gls{mmwave} frequencies provides higher resolution, while the higher penetration through many non-conducting materials compared to infrared waves enables the non-destructive inspection of hidden parts, making \gls{thz} an excellent candidate for this purpose. 
Moreover, high-throughput and low-latency \gls{thz} links can support real-time synchronization between real systems and their digital twin, enabling use cases requiring hardware-in-the-loop validation (e.g., virtual commissioning of industrial plants).

\section{Further considerations}\label{sec:further_cons}
This section provides further considerations on aspects related to the identified use cases, including propagation environments and deployment scenarios, other enabling technologies, and potential requirements and \glspl{kpi}.

\subsection{Propagation environments and deployment scenarios}
The identified use cases have been further analyzed and mapped to relevant propagation scenarios, as reported in \cite{etsigr001}.
These scenarios describe the physical propagation conditions that \gls{thz} communication systems will be subject to, consequently leading to the development of realistic channel models for the evaluation of technical solutions during the standardization phase.
For each of the 19 use cases, the report provides information on the target physical environments where those use cases are likely to be deployed (also reported in Tables \ref{tab:use_cases_1} and \ref{tab:use_cases_2}). Two groups have been defined: 13 indoor and 6 outdoor environments. 
The majority of use cases will be deployed in indoor environments, the most likely being living rooms, offices, and classrooms. Moreover, it shows that \gls{thz} communications will cover scenarios that are not currently supported by conventional wireless communications systems, such as intra-device, on-body, airplane and train cabins, and data centers.
Some use cases will be deployed in outdoor environments such as stadiums, open squares, urban canyons, and highways. In addition, \cite{etsigr001} provides a literature review on already available channel measurements and models revealing the physical environments that have not been already characterized. 
Few of them have been under-explored (on-body, medical facilities, living rooms, factories, classrooms, inter-machine, stadium and square environments), thus requiring further studies. This will be the subject of future work for the \gls{etsi} \gls{isg} \gls{thz}. 

Another important aspect to consider is the mobility of wireless terminals. In this regard, three different groups can be identified. The first group includes use cases with fixed nodes aiming to replace wired connections, such as wireless data centers, fixed wireless X-hauling, and intra-device communications. The second group includes use cases characterized by limited mobility, such as in-airplane or in-train cabin entertainment and local area collaboration between nearby devices. The third group includes use cases with mobile nodes, such as those related to robotic applications, wireless access for mobile users, and \gls{xr}. Given the high directionality that typically characterizes \gls{thz} links, the use cases belonging to the third group may require more advanced procedures for establishing and maintaining wireless connections compared to those with no or limited mobility. These considerations should be taken into account during the definition of technical requirements for \gls{thz} systems.

\subsection{Challenges and enablers}
Despite the tremendous potential, \gls{thz} communications present inherent challenges mainly related to the unique properties of signals propagating at these frequencies, including high sensitivity to blockage effects, limited propagation range, difficulties in channel measurements and modeling, pronounced near-field effects, high power consumption of \gls{rf} components. To solve these issues and exploit their full potential, the development of \gls{thz} wireless systems is expected to advance alongside other technologies, including:  

\subsubsection{\gls{ai} and \gls{ml}}
Unlike microwave bands, \gls{thz} signals can be easily blocked by physical objects such as walls, furniture, and even human bodies. Prediction algorithms such as recurrent neural networks can be leveraged to anticipate future blockage events and avoid link disruptions. 
Also, \gls{ml} techniques have the potential to streamline beam management operations, since traditional solutions lead to a huge signaling overhead due to the narrow size of \gls{thz} beams.  
Since performing channel measurements at such high frequencies is challenging, generative \gls{ai} approaches can be used to learn the statistical properties of \gls{thz} channels from a limited amount of measurement data and augment it with synthetic samples~\cite{ghosh2023thz}. Moreover, generalization techniques such as meta-learning can be adopted to adapt learned models to unseen environments/conditions.

\subsubsection{Advanced \gls{mimo} techniques}
The short wavelength at \gls{thz} allows for the employment of ultra-massive \gls{mimo} arrays, which can address challenges imposed by the wideband spectrum and severe propagation conditions. For example, novel \gls{thz} devices such as graphene-based plasmonic nano-antenna arrays can accommodate hundreds of antenna elements in a few millimeters. 
Advanced \gls{mimo} techniques such as \gls{los} or holographic \gls{mimo} can exploit the near-field effects that are more pronounced at \gls{thz} frequencies to shift from traditional beamforming, where the radiated energy is confined in an angular sector, to beam focusing, where the energy is directed to a specific point in space~\cite{singh2023utilization}. Unlike traditional massive \gls{mimo} schemes, these techniques enable spatial multiplexing even in \gls{los} conditions and increase spectral efficiency.
 
\subsubsection{\gls{ris} and advanced relays}
In indoor scenarios, \gls{thz} signals are easily blocked by walls, doors, and other building structures, therefore providing ubiquitous coverage can be an issue. In this regard, \glspl{ris} can mitigate the effect of blockage by creating additional propagation paths, enabling \gls{thz} communication and sensing in \gls{nlos} conditions. 
Transmissive \gls{ris} (T-RIS) can passively re-radiate the signal from an outdoor base station to indoor devices and mitigate losses due to \gls{o2i} penetration, which are particularly pronounced at high frequencies. Similarly, signal-transparent relays can avoid \gls{o2i} losses by translating \gls{thz} signals into optical signals and distributing them to indoor access points using the concept of radio-over-fiber. Moreover, relaying \gls{thz} signal through multiple hops using active or passive relays can extend their limited propagation range.

\subsubsection{Energy-efficient device technologies} Energy efficiency is a pre-requisite for \gls{thz} communications to achieve the desired link budget and limit power consumption. This is even more stringent for portable devices where battery life is a concern.  
In this regard, new materials such as graphene-based semiconductors are showing promising results for drastically improving energy efficiency compared to conventional electronic and photonic technologies. 

Other technologies recognized as particularly beneficial for supporting the identified use cases are indicated in Table~\ref{tab:use_cases_1}-\ref{tab:use_cases_2}, as reported in \cite{etsigr001}.

\subsection{Potential requirements and \glspl{kpi}}
The use cases that have been analyzed present diverse communication requirements which are summarized in Tab.~\ref{tab:kpi}. For example, some use cases may require latency below 0.5~ms, such as fixed wireless X-haul transport, cooperative mobile robots, real-time industrial control, and commissioning of industrial plants. Required data rates can exceed 1~Tbps, such as remote education, grand events with ultra-high throughput, and ultra-high throughput for indoor users. Use cases dealing with motion control and industrial applications, such as cooperative mobile robots, real-time industrial control, and commissioning of industrial plants may require extremely reliable operations. Some use cases may require a high connection density (up to 2 million devices/km$^2$) such as grand events with ultra-high throughput, ultra-high throughput for indoor users, and local area collaboration with mobility. Use cases such as interactive immersive \gls{xr} and simultaneous imaging, mapping, and localization have requirements related to sensing \glspl{kpi}, for example localization accuracy down to 0.5~cm. 
Power consumption and energy efficiency are considered key aspects for all identified use cases, especially for those involving portable devices with reduced form factors such as  
local area collaboration, cooperative mobile robots, \gls{xr}, grand events with ultra-high throughput, and ultra-high throughput for indoor users. 
In this regard, standardized metrics for comparing power consumption and energy efficiency across various system design are needed~\cite{ying2023waste}. 
The ETSI ISG THz will conduct further studies on this matter, and more detailed requirements will be presented in subsequent group reports.

\begin{table}
    \centering
    \caption{Summary of requirements related to the identified use cases.}
    \label{tab:kpi}
    \begin{tabular}{|c|c|}
        \hline
        End-to-end latency & $<$ 0.5 - 300~ms \\
        \hline
        Data rate & 10 Mbps - 1 Tbps \\
        \hline
        Reliability & 99,9 - 99,9999999 \% \\
        \hline
        Connection density & $<$ 2 million devices/km$^2$ \\
        \hline
        Localization accuracy & $>$ 0.5 cm \\
        \hline
    \end{tabular}
\end{table}

\section{Conclusion}\label{sec:concl}
6G communication systems will need to support the ever-increasing demands for high data rates as well as native integration of sensing functionalities. The distinctive features of the \gls{thz} frequency bands -- including high available bandwidth and sensing capabilities -- make \gls{thz} communications a suitable technology to support these demands. 
With the approval of the IMT-2030 framework~\cite{iturm2160}, the \gls{itu} agreed on the timeline for the development and standardization of 6G, where the period between 2024 and 2027 is dedicated to the definition of requirements and evaluation criteria. Within this time frame, the \gls{etsi} \gls{isg} \gls{thz} is setting the stage for future development and standardization of \gls{thz} wireless systems. Standardization activities related to technology specification are expected to start in 2027 and can make use of the outcomes of the \gls{etsi} \gls{isg} \gls{thz} as a starting point towards the inclusion of \gls{thz} communications into wireless standards.

This paper summarized the initial findings of the \gls{etsi} \gls{isg} \gls{thz}, including the identified use cases and frequency bands of interest for \gls{thz} communication systems, and provided an overview of standardization efforts focusing on this technology. 
The work described herein sets the basis for subsequent activities of the group, which will focus on channel and RF hardware modeling for \gls{thz} bands. 

\section*{Acknowledgement}
The authors would like to acknowledge the active members of the \gls{etsi} \gls{isg} \gls{thz} who contributed to the preparation of GR001 and GR002.

\bibliographystyle{ieeetr}
\bibliography{biblio.bib}

\end{document}

%% file: acronyms.tex
\newacronym{thz}{THz}{terahertz}
\newacronym{etsi}{ETSI}{European Telecommunications Standards Institute}
\newacronym{isg}{ISG}{Industry Specification Group}
\newacronym{itu}{ITU}{International Telecommunication Union}
\newacronym{wrc}{WRC} {World Radiocommunication Conference}
\newacronym{eess}{EESS}{Earth Exploration Satellite Services}
\newacronym{ras}{RAS}{Radio Astronomy Services}
\newacronym{ieee}{IEEE}{Institute of Electrical and Electronics Engineers}
\newacronym{3gpp}{3GPP}{3rd Generation Partnership Project}
\newacronym{gr}{GR}{Group Report}
\newacronym{xr}{XR}{eXtended Reality}
\newacronym{vr}{VR}{Augmented Reality}
\newacronym{ar}{AR}{Virtual Reality}
\newacronym{mr}{MR}{Mixed Reality}
\newacronym{d2d}{D2D}{Device-to-Device}
\newacronym{iot}{IoT}{Internet of Things}
\newacronym{dt}{DT}{Digital Twin}
\newacronym{kpi}{KPI}{Key Performance Indicator}
\newacronym{mmwave}{mmWave}{millimeter wave}
\newacronym{ai}{AI}{Artificial Intelligence}
\newacronym{mimo}{MIMO}{Multiple Input Multiple Output}
\newacronym{los}{LOS}{Line Of Sight}
\newacronym{nlos}{NLOS}{Non Line Of Sight}
\newacronym{ris}{RIS}{Reconfigurable Intelligent Surface}
\newacronym{em}{EM}{electromagnetic}
\newacronym{isac}{ISAC}{Integrated Sensing and Communication}
\newacronym{o2i}{O2I}{Outdoor-to-Indoor}
\newacronym{ue}{UE}{User Equipment}
\newacronym{llm}{LLM}{Large Language Model}
\newacronym{ml}{ML}{Machine Learning}
\newacronym{6g}{6G}{sixth generation}
\newacronym{emf}{EMF}{Electromagnetic Field}
\newacronym{rf}{RF}{Radio Frequency}